\documentclass[12pt, a4paper]{article}
\usepackage[font=footnotesize]{caption}
\usepackage[utf8]{inputenc}
\usepackage{graphicx}
\usepackage{authblk}
\usepackage{booktabs}
\usepackage{verbatim}
\usepackage{url}
\usepackage{caption}
\usepackage{subcaption}
\usepackage{indentfirst}
\usepackage{url}
\usepackage{amssymb}
\usepackage{amsmath}
\title{Distance-Based Hierarchical Cutting of Complex Networks with Non-Preferential and Preferential Choice of Seeds}
\author{Alexandre Benatti$^1$ \\ Luciano da F. Costa$^2$}

\affil{
$^1$Institute of Mathematics and Statistics - DCC \\
University of S\~ao Paulo \\
Rua do Mat\~ao, 1010, \\ S\~ao Paulo, SP 05508-090 Brazil 
\\ \vspace{0.5cm}
$^2$S\~ao Carlos Institute of Physics - DFCM \\
University of S\~ao Paulo \\
Av. Trabalhador S\~ao-carlense, 400, \\ S\~ao Carlos, SP 13566-590 Brazil
}

\date{\emph{15th March, 2024}}

\begin{document}

\maketitle

\begin{abstract}
Graphs and complex networks can be successively separated into connected components associated to respective seed nodes, therefore establishing a respective hierarchical organization. In the present work, we study the properties of the hierarchical structure implied by distance-based cutting of Erd\H{o}s-R\'enyi, Barab\'asi-Albert, and a specific geometric network. Two main situations are considered regarding the choice of the seeds: non-preferential and preferential to the respective node degree. Among the obtained findings, we have the tendency of geometrical networks yielding more balanced pairs of connected components along the network progressive separation, presenting little chaining effects, followed by the Erd\H{o}s-R\'enyi and Barab\'asi-Albert types of networks. The choice of seeds preferential to the node degree tended to enhance the balance of the connected components in the case of the geometrical networks.
\end{abstract}

\section{Introduction}\label{sec:introduction}

The way in which graphs and complex networks react to topological changes can provide interesting information about their respective properties, including resilience (e.g.~\cite{albert2000error,motter2002cascade,crucitti2004error,miorandi2010k,gao2016universal,artime2024,shao2015percolation}). In this work, we approach the situation in which graphs are progressively separated into new graphs as a consequence of their nodes being associated to distinct seed, or reference, nodes. Interestingly, in a manner similar to that observed in~\cite{benatti2024hierarchical} respectively to graph cutting performed by random walks, the successive separation of graphs among involved seed nodes also establish a respective \emph{hierarchy} between the obtained connected components, which can be effectively visualized in terms of respective \emph{dendrograms}. The hierarchical structure of complex networks consists of an area that has motivated continuing attention in the respective literature (e.g.~\cite{sales2007extracting,clauset2008hierarchical,benatti2023recovering,barabasi2003hierarchical,corominas2013origins,meyerhenke2017,benatti2024hierarchical}).

\begin{figure}
  \centering
     \includegraphics[width=.99 \textwidth]{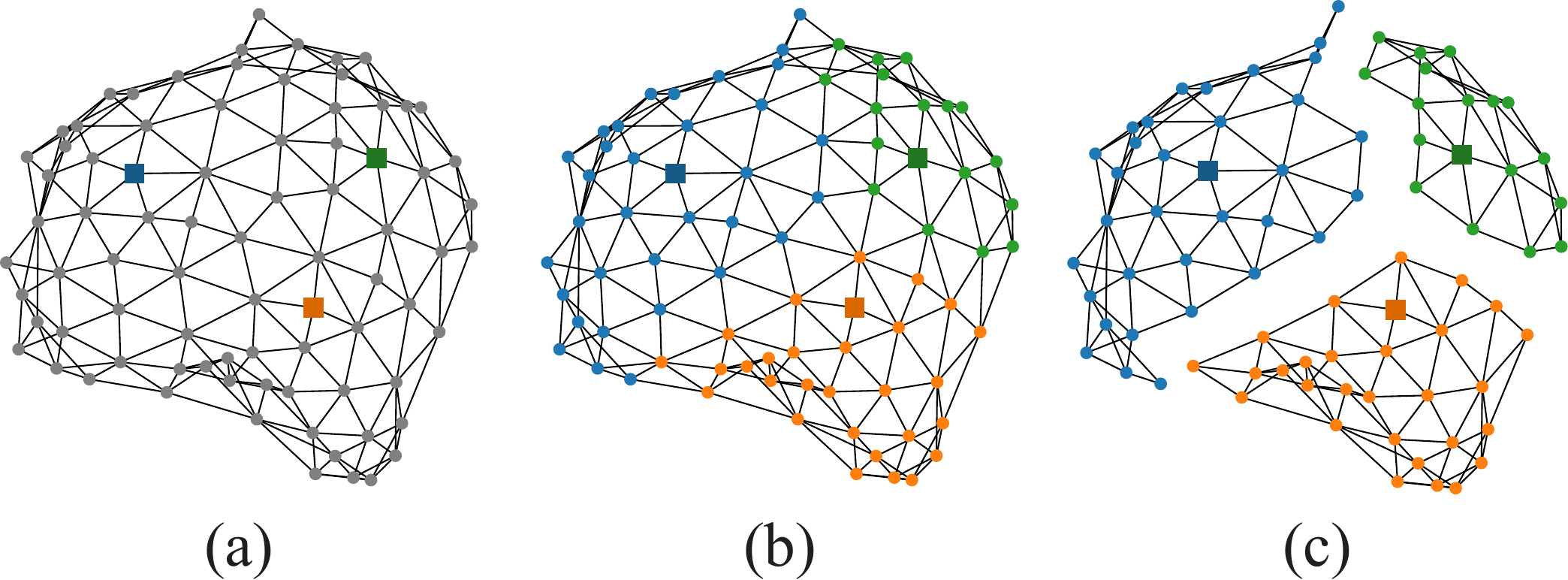}
   \caption{An original graph (a) with three seed nodes identified as square nodes in green, blue, and orange. The same graph partitioned (b), in terms of the shortest topological distances, between the three seed nodes. Nodes that have the same distance to two or more seeds are assigned randomly to one of those seeds, in a manner than can be understood as being analog to the concept of Dirichlet tessellation (or Voronoi diagrams, e.g.~\cite{riedinger1988delaunay}). Once partitioned, the graph can be separated (c) into connected components respective to each of the seeds by removing the edges extending between nodes belonging to different partitions. This basic separation approach can then be repeated, yielding a respective hierarchy.}\label{fig:example}
\end{figure}

Given a complex network (or graph) $\Gamma$ with $N$ nodes, as well as a set $S$ containing $M$ of these nodes (without repetition), which are henceforth called \emph{seed nodes}, it is possible to \emph{partition} (e.g.~\cite{meyerhenke2017}) $\Gamma$ into $M$ respective subgraphs defined by the nodes that are closer to each of the respective seeds. These subgraphs can then be separated, yielding $M$ independent connected components which are, themselves, new graphs. Figure~\ref{fig:example} illustrates the above described procedure respectively to a simple graph.

In addition to its intrinsic theoretical interest, the above described operation, which is henceforth called \emph{hierarchical cutting} of a complex network, potentially underlies some real-world situations of interest. These include the establishment of countries around respective reference capital cities, which can be subdivided into states and respective capitals, municipalities, and city regions (boroughs). Similarly, knowledge networks (e.g.~\cite{da2006learning,ferraz2018,lu2022}) can also be successively subdivided into subgraphs established respectively to specific sets of seed nodes (main topics).

From the above context, the present work can be understood as providing subsidies not only for modeling these types of hierarchical networks but also for characterizing their specific topological properties which, in the present work (as in~\cite{benatti2024hierarchical}), is performed while paying special attention to size and balance of each of the obtained new pair of connected components.

These properties are of particular interest as they directly characterize how the components are successively separated, also providing an indication of the respective resilience to the cuts. For instance, a previous study~\cite{benatti2024hierarchical} has indicated that the sectioning of complex networks of the ER and BA types performed by random walks tend to yield chained dendrograms (e.g.~\cite{duda1973,costa2000shape,corominas2013origins,tokuda2022revisiting}), characterized by successive separations into a large group and another group often involving only one node. This type of chained cutting can be understood to be strongly unbalanced, implying the tendency of a large component to be found at most hierarchical levels.

Chained cutting tends to allow relatively large components (though appearing in unbalanced manner) extending along several cutting hierarchies. On the other hand, more balanced cuttings involving pairs of connected components with not only similar but also relatively large sizes, tend to require fewer hierarchical levels before the complete separation of the original network. It should be observed that chained or more balanced cuttings can be both intrinsically desirable or undesirable, depending on specific problems and respective objectives. 

One aspect of specific interest developed in the present work concerns the study of how the adoption of a distance-based methodology for cutting graphs and complex networks can affect the sizes and balance of the obtained components. Because the intrinsic topology of each type of complex network model can be expected to influence the successive network separation in specific manners, we shall consider three distinct models of networks in the reported studies, namely Erdős–Rényi (ER)~\cite{erdos1959random}, Barabási–Albert (BA)~\cite{barabasi1999BA}, and a specific Geometric Graph (GEO) -- e.g.~\cite{riedinger1988delaunay,benatti2023simple}. As in~\cite{benatti2024hierarchical}, we resource to scatterplots in the component size coordinate space $\left(n_1, n_2 \right)$ in order to better understand the size and balance of the connected components obtained along the hierarchical cutting procedure.

In addition to choosing the seeds uniformly among the existing nodes, we also study the effect of choosing them preferentially to the node degree. Even though ER and GEO networks tend to have nodes with similar degree, statistical fluctuations can generate a few nodes with degree substantially larger and smaller than the respective average. Thus, the use of non-preferential and preferential choice of seeds can be expected to have influence in the hierarchical cuttings obtained not only for the BA networks, which are intrinsically degree-skewed but also on the considered ER and GEO structures.

This work starts by describing the main concepts and methods and follows by presenting the obtained results and the respective discussion.

\section{Concepts and Methods}\label{sec:methods}

Given a network (or graph) $\Gamma$ with $N$ nodes (the network \emph{size}), it is possible to calculate the \emph{topological distance} $\delta_{ij}$ between each pair of nodes $(i,j)$, which can be organized as a respective \emph{distance matrix} $D$.

Now, let us consider a set $S$ of $M$ nodes sampled from the $N$ original nodes without repetition. These nodes will be henceforth called \emph{seed nodes} or, simply, \emph{seeds}.

In case a specific node $k$ is selected from the set $S$, the network nodes that are closest to $k$ than to any of the other seed nodes can be identified, which can be understood as a \emph{region of influence} of the node $k$ in the network $\Gamma$. Nodes that are equally distant to more than one of the seeds are randomly assigned among those nodes. The identification of the regions of influence respective to each of the seed nodes therefore establishes a \emph{partitioning} of the original graph into $M$ respective subgraphs.

Once a partitioning of the original network is obtained respectively to the $M$ seeds, it becomes possible to \emph{separate} or \emph{disconnect} these subgraphs, which can be done by removing every edge that interconnects nodes belonging to distinct regions of influence.

\begin{figure}
  \centering
     \includegraphics[width=.9 \textwidth]{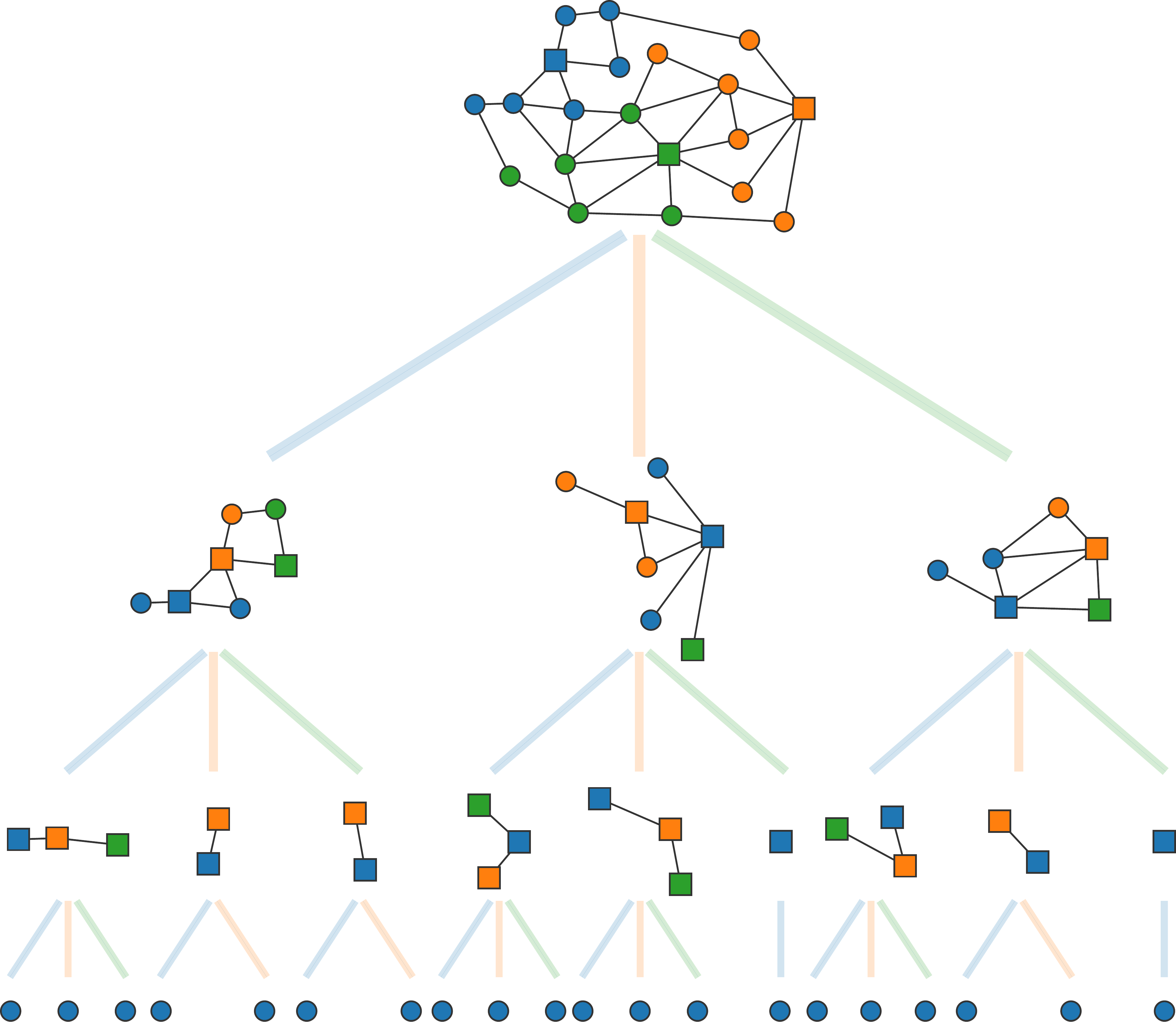}
   \caption{Illustration of the recurrent application of the distance-based cutting of networks respectively to $M=3$ seeds, which are shown as square nodes.}\label{fig:branch}
\end{figure}

\begin{figure}
  \centering
     \includegraphics[width=.55 \textwidth]{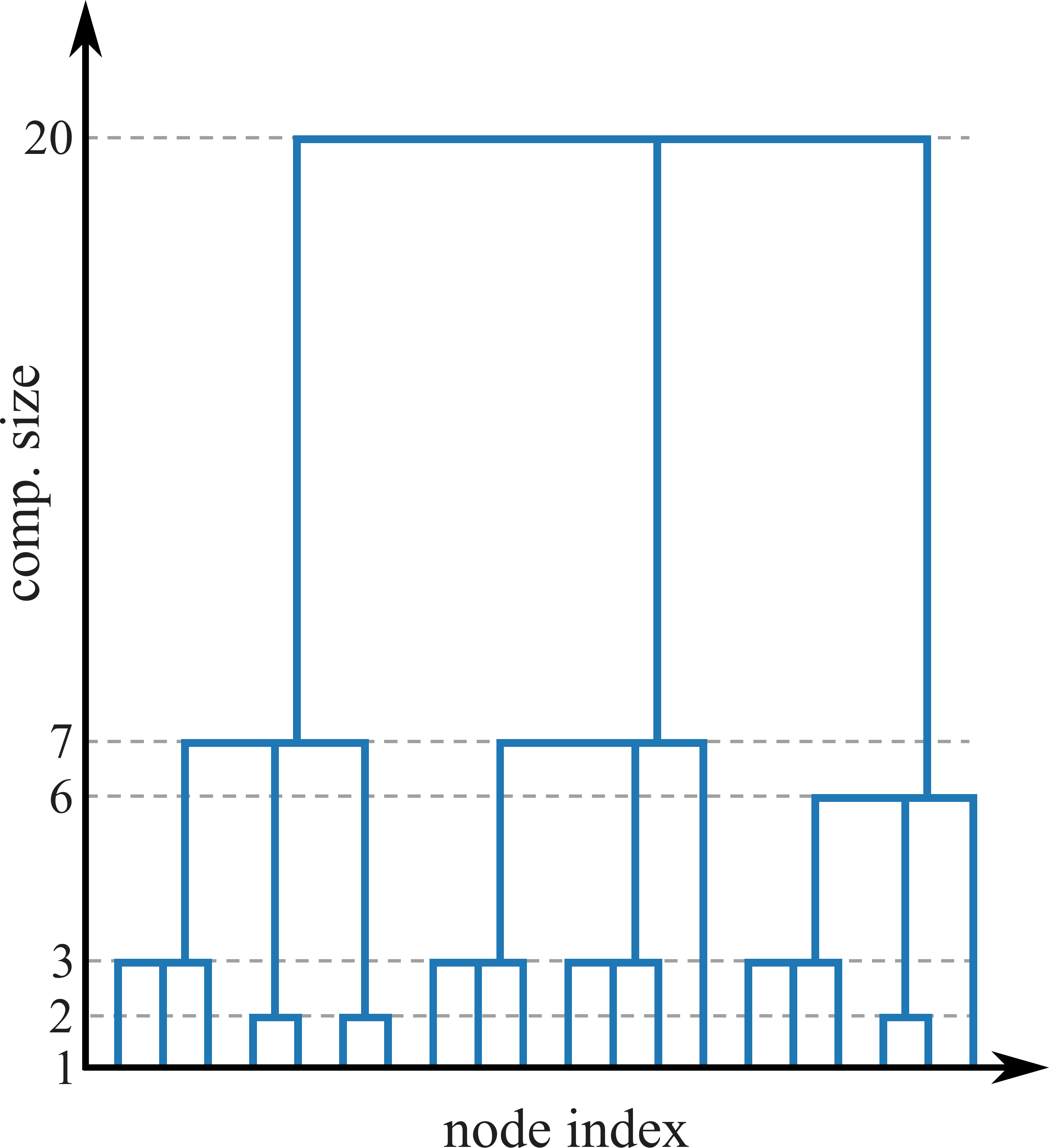}
   \caption{Dendrogram obtained from the distance-based cutting illustrated in Fig.~\ref{fig:branch_dendrogram}.}\label{fig:branch_dendrogram}
\end{figure}

The above described separation of a graph into $M$ connected components, which is illustrated in Figure~\ref{fig:branch} for $M=3$, can then be repeated in recurrent manner, with $M$ seed nodes being chosen within each of the obtained connected components. The recurrence ends when all connected components have a size of 1. The obtained successive cuttings establish a respective \emph{hierarchy} among the obtained components, as illustrated in Figure~\ref{fig:branch_dendrogram}, 
which can also be represented as a respective \emph{dendrogram}.

It is interesting to assign a tuple $\left(n_1, n_2, \ldots, n_M \right)$ to each branching, where $n_i$ corresponds to the size of the respective components. It is also assumed that $n1 \leq n_2 \leq \ldots \leq n_M$.

In the particular case of $M=2$, the graphs will always be separated into two respective connected components, yielding a respective dendrogram which is also a \emph{binary tree}. In this case, it is possible to represent all possible separations in terms of their respective tuples $\left(n_1, n_2 \right)$, which can be visualized in terms of a respective \emph{scatterplot}.

\begin{figure}
  \centering
     \includegraphics[width=.6 \textwidth]{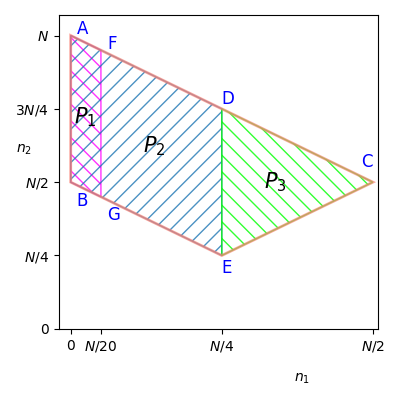}
   \caption{The region of possibly observable coordinates
   $\left(n_1, n_2\right)$, assuming the components to have al least size $N/2$, is defined by four vertices $A, B, C,$ and $E$. The most balanced situation, characterized by the two resulting components having the same size, corresponds to the point $C$. Henceforth, the region within the bounding polygon is separated into two sub-regions delimitated by the polygons $ABDE$ and $CDE$, with the latter being associated to more balanced pairs of connected components which are also relatively large. The total density of observations resulting within these two regions are henceforth expressed as $P_2$ and $P_3$. A portion of the region $ABDE$, namely that comprised within $AFGB$, is also considered to correspond to chained pairs of components, leading to a respective probability $P_1$. }\label{fig:Geometry}
\end{figure}

Figure~\ref{fig:Geometry} depicts the polygonal region defining the coordinates $\left(n_1, n_2 \right)$ of possible connected components in the case $M=2$. It is henceforth assumed that only connected components with a size larger or equal to $N/2$ are taken into account while preparing this type of scatterplot and regions.

The line $AB$ indicates that $n_1,n_2 > 0$, and the line segment $BE$ is implied by $n_1 + n_2 \geq N/2$. Similarly, the bounding line $AC$ corresponds to the cases in which $n_1+n_2=N$, being therefore respective to the first separation of the original network. The line $DE$ establishes a separation of the bounded region into respectively to the two following configurations: (1) $ADEB$ pairs of component sizes that are relatively small and less balanced; and (2) $DCE$ pairs of component sizes that are relatively large and more balanced. 

Maximum balance is obtained for the cases resulting in point $C$. The total density (probability) at the two above regions is henceforth indicated as $P_2$ and $P_3$, with $P_2+P_3=1$. A portion of the region $ADEB$, namely $AFGB$, is understood to correspond to chained pairs of connected components, being characterized by $n_1 \ll n_2$. Observe that the extension of region $AFGB$ along the horizontal axis is equal to $N/10$. The probability of having chained pairs of components can thus be obtained by integrating the density within the region $ADEB$, yielding respective probability $P_3$. Observe the existence of overlap between the regions $ADEB$ and $AFGB$, so that $P_1 + P_2 + P_3 \geq 1$.

It is interesting to observe that the above described diagram can also be constructed considering normalized values of the size of pairs of connected component sizes, so that they yield respective fractions $\left(f_1,f_2 \right)$ --- with $f_1=n_1/n$, $f_2=n_2/n$ and $f_1+f_2=1$, of the size $n$ of the connected component from which they originated.

Table~\ref{tab:probs} summarizes the three probabilities considered in this work as a means to characterized the balance and chained of the pairs of connected components obtained along distance-based cutting of complex networks.

\begin{table}
\centering
\caption{The three probabilities adopted for characterizing balance and chaining of pairs of connected components obtained along distance-based cutting of complex networks. In all cases $n_1 \leq n_2$, $P_2+P_3=1$, and $P_1 \leq P_2$.}\label{tab:probs}
\begin{tabular}{|c|c|c|}
\hline 
\emph{Prob.} & \emph{Region} & \emph{Meaning} \\
\hline
$P_1$ & $AFGB$ & Prob. of chained pairs ($n_1 < N/20$, $n_1 \ll n_2$) \\ \hline
$P_2$ & $ADEB$ & Prob. of non-balanced pairs ($n_1 < N/4$) \\ \hline
$P_3$ & $DCE$ & Prob. of nearly balanced pairs ($n_1 \geq N/4$) \\ \hline
\end{tabular}
\end{table}

\section{Results and Discussion}\label{sec:results}

Experiments have been performed considering ER, BA, and GEO networks with $N=100$ nodes and average degree $\left< k \right> \approx 5.7$. The results obtained respectively to non-preferential and preferential choice of the seeds (respectively to the node degree) are reported in the following sub-sections.

\subsection{Non-Preferential Choice of Seeds}

\begin{figure}
  \centering
     \includegraphics[width=.325 \textwidth]{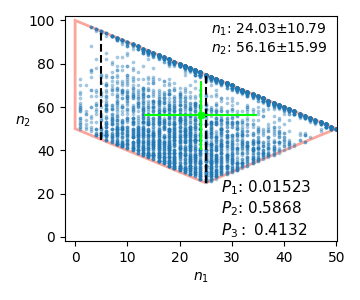}
     \includegraphics[width=.325 \textwidth]{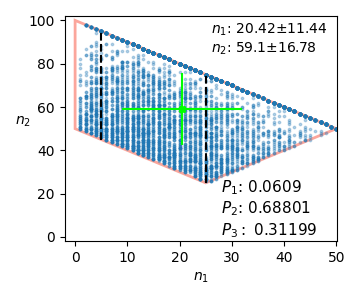}
     \includegraphics[width=.325 \textwidth]{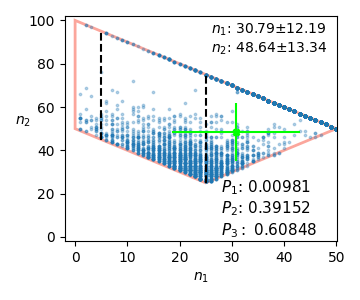}
     (a) \hspace{3.7cm} (b) \hspace{3.7cm} (c)
   \caption{Scatterplots of component sizes $\left(n_1, n_2 \right)$ obtained for 2000 networks of ER (a), BA(b), and GEO (c) types considering the uniform choice of the seed nodes. The points on the line segment $AC$ correspond to the very first separation of the original network ($n_1+n_2=N$). The green cross-hair indicates the average $\pm$ standard deviation of the values of $n_1$ and $n_2$. The networks of GEO type resulted in the most balanced components, indicated by the concentration of cases near point E. At the same time, the BA network tended to yield the less balanced connected components, with the observed coordinates covering most of the bounding polygon.}\label{fig:scatter}
\end{figure}

At each separation stage, the $M$ nodes are chosen uniformly among the existing nodes. Figure~\ref{fig:scatter} presents the scatterplots $\left(n_1, n_2 \right)$ obtained for 2000 networks of the three considered types.

Interestingly, the GEO networks yielded a concentration of points that mostly surround the point $E$. At the same time, the observed initial separations, lying on line $AC$, shifted to the right-hand side of the figure, also indicating an enhanced balance of component sizes. The less balanced connected components have been obtained for the BA networks, with the observed points extending widely within the bounding region. The ER networks led to intermediate results.

The enhanced balance of component sizes observed for the GEO case are related to the intrinsic topological uniformity of this type of networks, as well as for the fact that these are also not small world, having nodes conforming to a two-dimensional space.

\begin{figure}
  \centering
     \includegraphics[width=1. \textwidth]{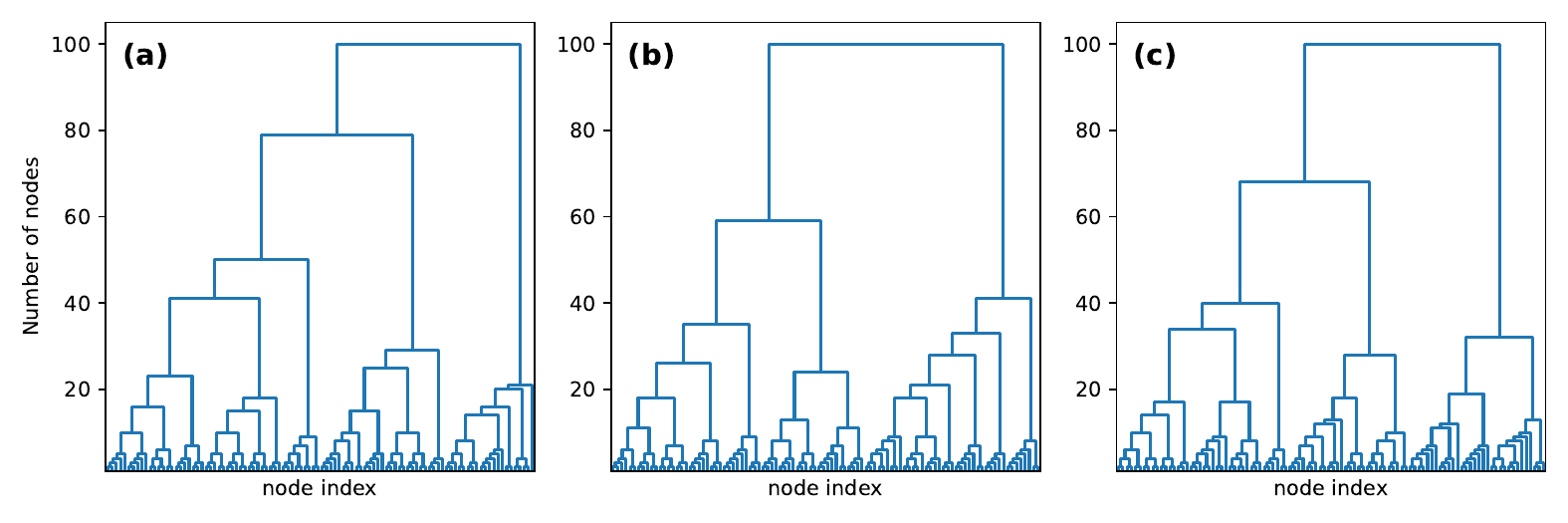}
     \includegraphics[width=1. \textwidth]{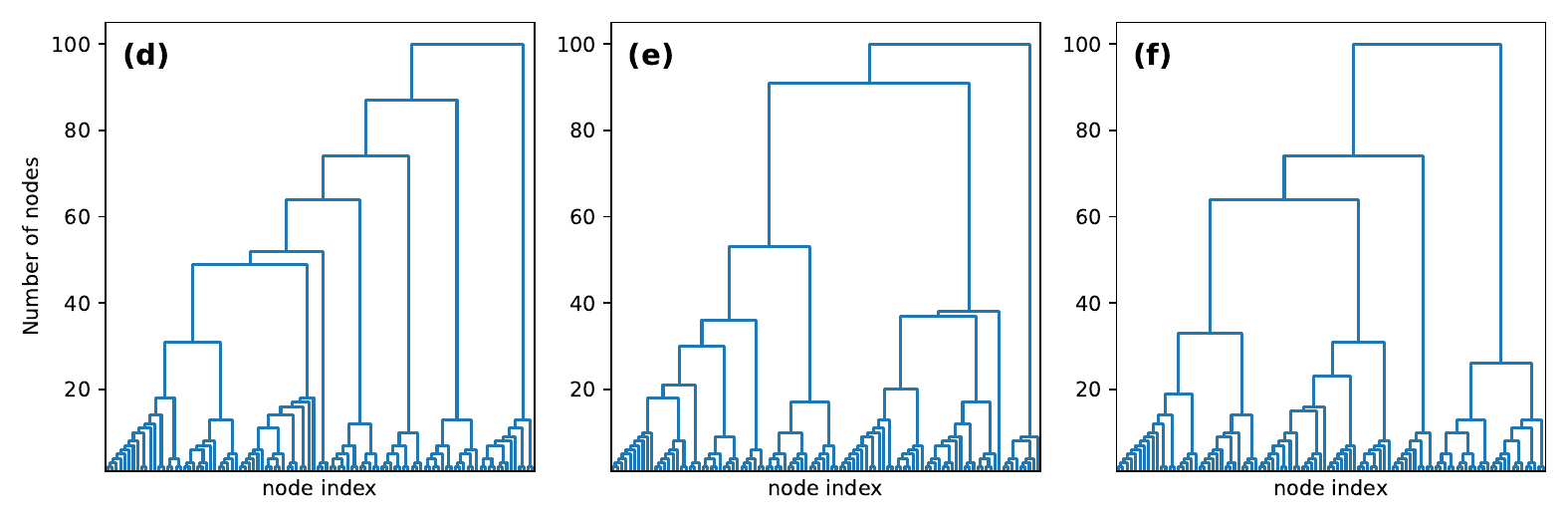}
     \includegraphics[width=1. \textwidth]{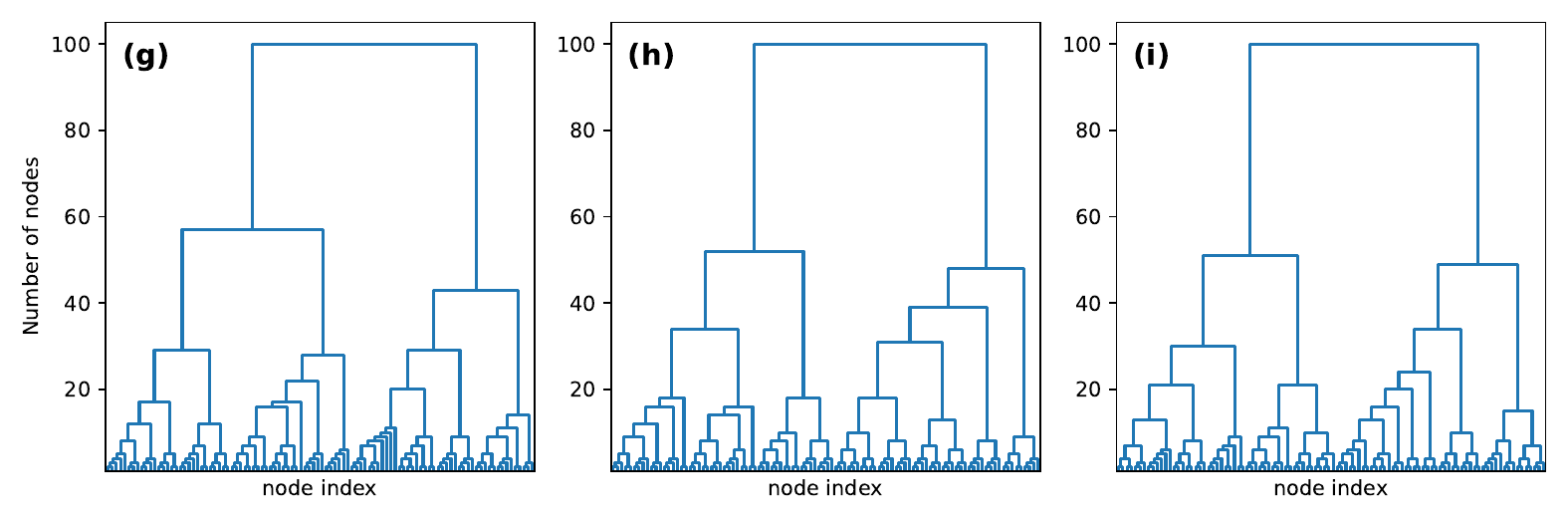}
   \caption{Examples of dendrograms obtained for the ER (a--c), BA (d--f), and GEO (g--i) types of networks subjected to distance-based hierarchical cutting, considering a uniform choice of seeds. The vertical axis corresponds to the \emph{size} of the connected components, while the horizontal axes refers to the labels of the network nodes. }\label{fig:dendrogram}
\end{figure}

Figure~\ref{fig:dendrogram} provides three illustrations of dendrograms obtained for each of the three network types considered in the present work. Each line corresponds to a network type.

The dendrograms (g--i) corresponding to GEO networks resulted substantially more balanced, with little \emph{chaining}, thus confirming the previously reported and discussed results. At the same time, the BA networks led to the more \emph{chained} dendrograms, which are consequently particularly unbalanced, with one of the connected components tending to be substantially larger than the other component in each respective pair. The ER networks resulted with intermediate properties between those obtained for the GEO and BA networks.

\subsection{Preferential Choice of Seeds}

In this sub-section we repeat all the previous experiments, but now choosing the seeds proportionally to the node degree.

\begin{figure}
  \centering
     \includegraphics[width=.325 \textwidth]{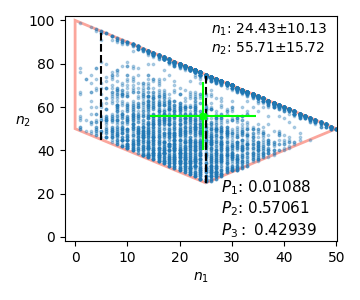}
     \includegraphics[width=.325 \textwidth]{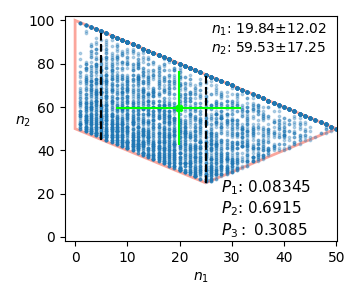}
     \includegraphics[width=.325 \textwidth]{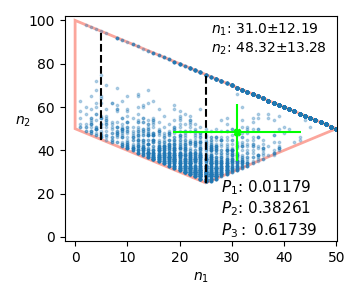}
     (a) \hspace{3.7cm} (b) \hspace{3.7cm} (c)
   \caption{Scatterplots of component sizes $\left(n_1, n_2 \right)$ obtained for 2000 networks of ER (a), BA(b), and GEO (c) types considering the choice of the seeds as being preferential to the node degree. Interestingly, this alternative strategy led to even further enhanced balance of the connected components obtained in the GEO networks.}\label{fig:scatter_pref}
\end{figure}

Figure~\ref{fig:scatter_pref} presents the scatterplots obtained for the ER, BA, and GEO networks.

Interestingly, in the case of GEO networks, the preferential choice of seed nodes led to even more balanced and relatively large connected components, as also reflected on the larger value of $P_2$ obtained for this type of cutting. This effect could be related to the fact that the choice of seed nodes with a large degree in GEO networks tended to provide a more effective partitioning of the associated regions characterized by less intense statistical dispersion while determining the respective regions of influence.

Also of particular interest, while the preferential approach led to minor differences in the case of the ER networks, in the case of BA networks it yielded connected components that are less balanced than those obtained in the case of non-preferential choice of seeds. That is possibly a consequence of the fact that the latter type of choice tended to enhance the intense original degree skew already present in the original BA networks.

\begin{figure}
  \centering
     \includegraphics[width=1. \textwidth]{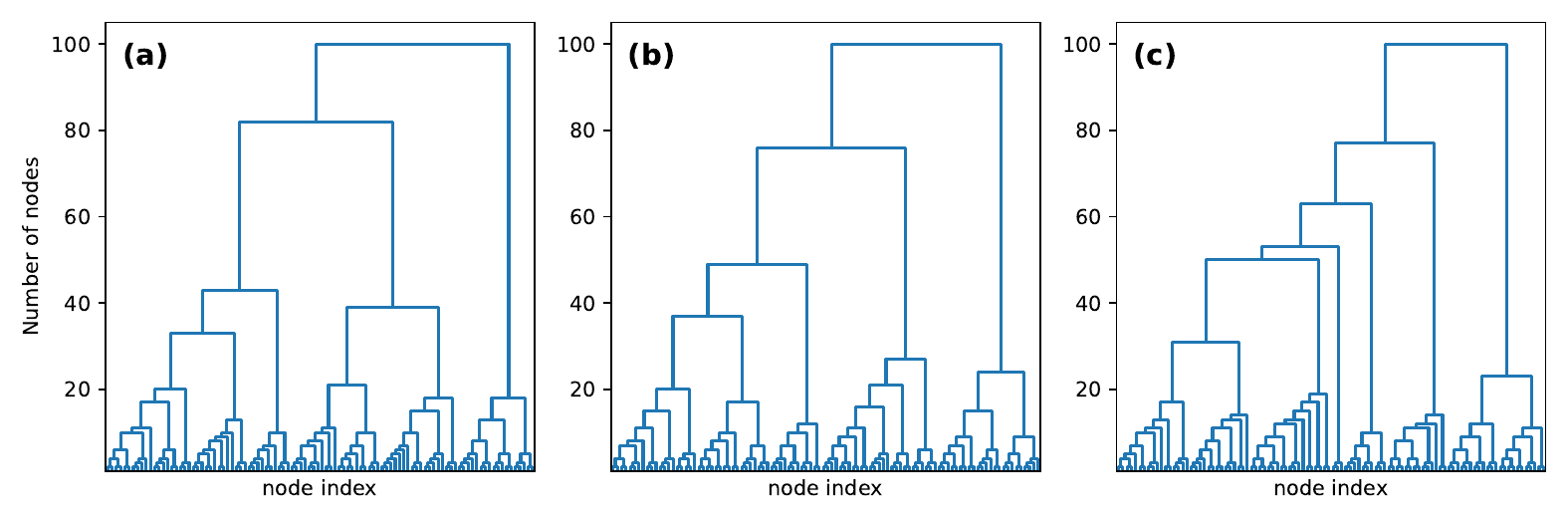}
     \includegraphics[width=1. \textwidth]{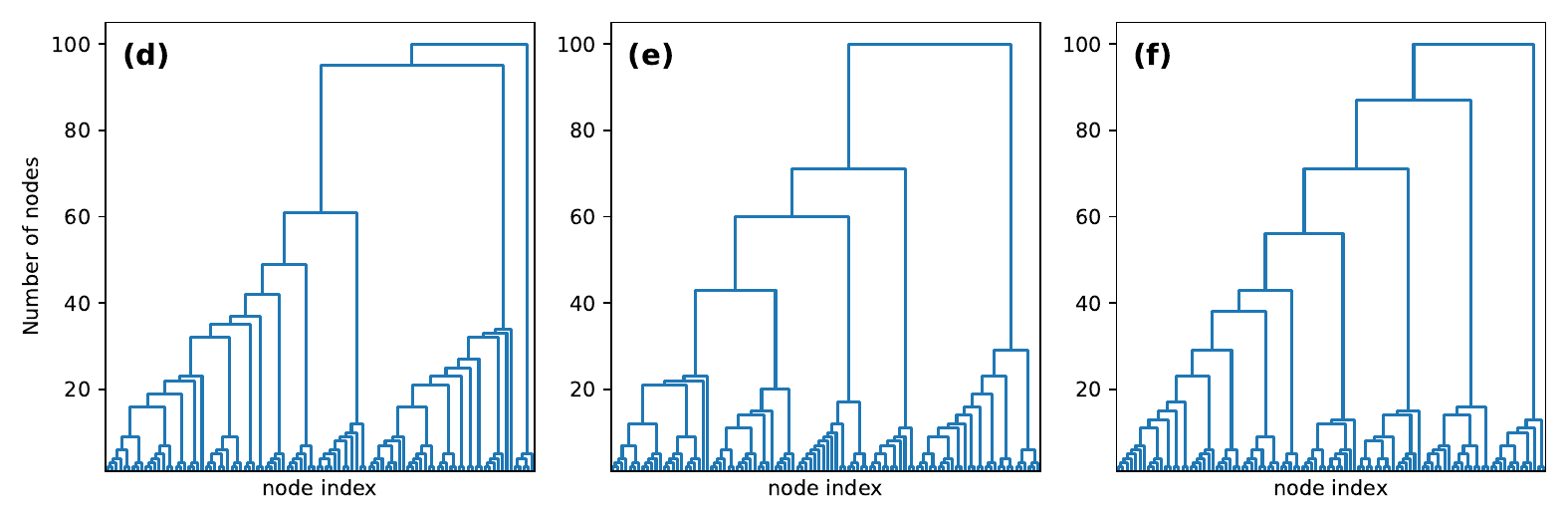}
     \includegraphics[width=1. \textwidth]{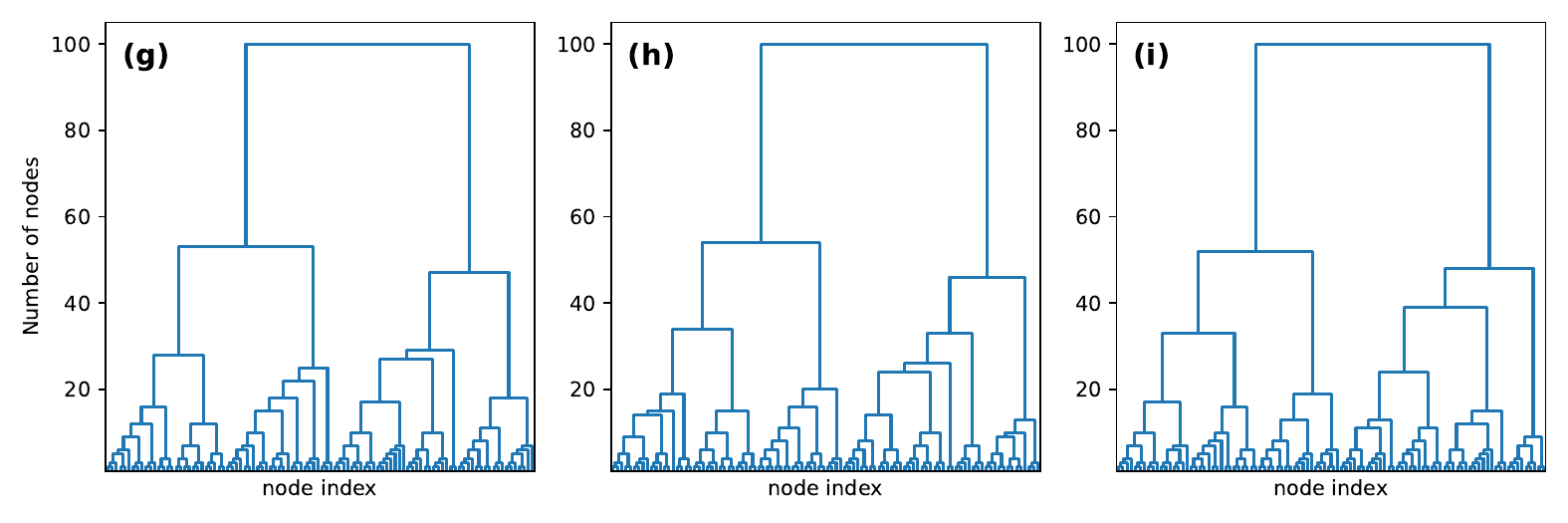}
   \caption{Examples of dendrograms obtained for the ER (a--c), BA (d--f), and GEO (g--i) types of networks subjected to distance-based hierarchical cutting, with the choice of seeds being preferential to the node degree. Interestingly, intensely chained dendrograms have been obtained for the distance-based cutting of BA networks (d--f) when the seeds are chosen proportionally to the node degree.}\label{fig:dendrogram_pref}
\end{figure}

Further interesting insights about the effects of preferential choice of the seeds can be observed in the example dendrograms in Figure~\ref{fig:dendrogram_pref}.

While a relatively minor effect of preferential choice can be observed respectively to the ER and GEO cases, the BA dendrograms resulted in substantially more chained. This could be an effect of the scale-free structure of these types of networks, which are therefore characterized by the tendency to have hubs. More specifically, a seed corresponding to a hub node will tend to define a large respective region of influence, and therefore, of connected component size, with this tendency recurring along the topological scales as a consequence of the scale-free nature of BA networks. 

As indicated by the probabilities $P_1$ and $P_2$, the consideration of preferential choice of seeds in the ER and GEO network cases also led to slightly enhanced balanced pairs of connected components.

\section{Distance-Based Cutting with Three Seeds}

The experimental results previously reported in this work were characterized by $M=2$, leading to cutting hierarchies that correspond to binary trees. The present section provides some illustration of cases considering $M=3$ seeds. Because the diagrams as described in Figure~\ref{fig:Geometry} are not possible in this case, we henceforth limit our attention to dendrogram visualizations of the obtained hierarchies.

\begin{figure}
  \centering
     \includegraphics[width=1. \textwidth]{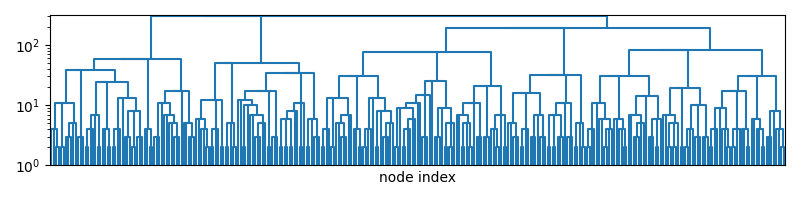} \\ (a) \\
     \includegraphics[width=1. \textwidth]{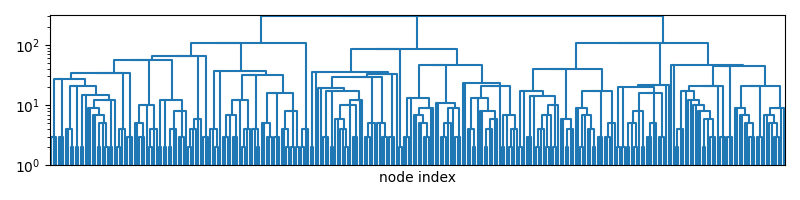} \\ (b) \\
     \includegraphics[width=1. \textwidth]{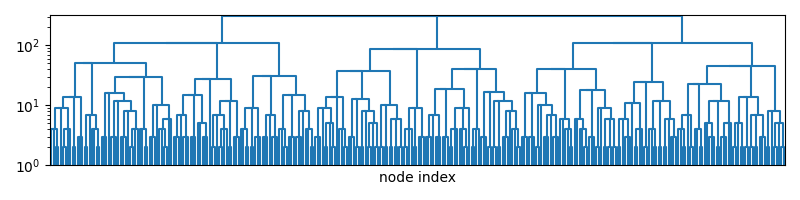} \\ (c)
   \caption{Examples of dendrograms obtained for the ER (a), BA (b), and GEO (c) types of networks subjected to distance-based hierarchical cutting, with the choice of seeds being non-preferential to the node degree. Similar tendencies can be observed respectively to the case $M=2$, with the GEO network leading to the more balanced dendrograms, followed by the ER and BA cases.}\label{fig:dendrogram_3}
\end{figure}

Figure~\ref{fig:dendrogram_3} illustrates dendrograms obtained by the distance-based hierarchical cutting of ER, BA, and GEO models, with the seeds being chosen uniformly (non-preferentially to the node degree).  Networks of size $N=300$ have been considered in order to obtain a larger number of hierarchical levels. The horizontal axes are shown in logarithm scale for the sake of better visualization, given that the size of the connected components decreases much more steadily along the hierarchies as in the case $M=2$.

Interestingly, the same tendencies observed for the case $M=2$ can be observed from Figure~\ref{fig:dendrogram_3}. More specifically, the GEO model led to the more balanced dendrograms, followed by the ER and BA cases. The latter type of networks were again characterized by intense chaining and unbalance along the respective hierarchies.

\section{Concluding Remarks}

The topological and dynamical properties of graphs and complex networks have been extensively studied from the point of view of respective measurements (e.g. \cite{newman2018networks,barabasi2013network,costa2007characterization}). At the same time, analyzing how graphs and networks undergo successive, hierarchical cutting under some specific type of action, as performed by random walks or distance-based node assignment among seeds, has the potential for providing further insights about the intrinsic topological and dynamic properties of graphs and networks.

In the present work, we approached the modeling and characterization of distance-based hierarchical cutting of three types of complex networks, namely ER, BA, as well as a specific geometric model GEO. Scatterplots of the sizes of pairs of obtained connected components have been employed as a means to visualize and analyze the effects of distance-based hierarchical cutting of the three considered types of network models while taking into account two ways of choosing the seeds (non-preferential and preferential to the node degree).

The obtained results indicate that, at least for the considered types and configurations of networks, the distance-based cutting of the specific type of GEO networks led to the most balanced and relatively large sizes of connected components, with dendrograms characterized by little chaining. At the same time, the BA networks yielded less-balanced pairs of connected components, with dendrograms incorporating intense chaining.

Interesting results have been identified also regarding the effect of performing distance-based hierarchical cutting of networks while choosing the respective seeds preferentially to the node degree. More specifically, this type of cutting led to even relatively larger and more balanced pairs of connected components appearing along the cutting of GEO networks. Also of interest, the preferential approach led to less balanced cuttings in the case of BA networks.

In addition to their many theoretical implications, the obtained results are also potentially important in real-world situations. For instance, they show that it could be particularly difficult to separate scale-free networks into sub-components with similar, balanced sizes by using distance-related approaches. At the same time, even random distance-based separation of geometrical networks tends to yield more balanced and relatively large connected components. Thus, in case knowledge networks have a scale-free structure, the obtention of respective balanced ontologies would tend to require specific heuristics and methodologies to be developed and applied.

A preliminary consideration of $M=3$ indicated that the effects observed for $M=2$ tend to be conserved.  More specifically, more balanced dendrograms were obtained for the GEO networks, followed by the ER and BA cases.

It is also interesting to contrast the here reported results with the previous study of hierarchical cutting of networks by random-walks~\cite{benatti2024hierarchical}. Interestingly, the distance-based cutting methodology yielded substantially less chained dendrograms, with relatively large and more balanced connected components obtained along the cutting dynamics. Contrariwise, the cutting of complex networks by random walks tends to result in highly chained hierarchies, with the geometrical model tending to allow moderately more balanced and relative large pairs of connected components. This result therefore suggests a fundamental difference in the distance-based and random walk-based hierarchical cutting of complex networks.

The reported concepts, methods, and results pave the way to a number of related further research, including the consideration of other sizes, configurations, and types of networks such as those involving modular structure, variations of the cutting strategy (e.g.~along the spanning tree associated to the complex networks), as well as the study of the impact of the cuttings on other networks properties such as modularity, overall distances, assortativity, and clustering, among other possibilities.  Another interesting possibility is to generalize the described two-dimensional diagram for hierarchical cuttings performed with more than two seeds.

\section*{Acknowledgments}
Alexandre Benatti thanks MCTI PPI-SOFTEX (TIC 13 DOU 01245.010\\222/2022-44). Luciano da F. Costa thanks CNPq (grant no.~307085/2018-0) and FAPESP (grant 2022/15304-4).

\bibliography{ref}
\bibliographystyle{unsrt}

\end{document}